\documentclass[12pt,preprint]{aastex}

\usepackage{natbib} % in-line citation package (see explanations in natbib.sty)
\bibpunct[,]{(}{)}{;}{a}{}{,} % ApJ in-line citation style

\def\gta{\ifmmode {\mathbin{\lower 3pt\hbox   %> or of order
    {$\,\rlap{\raise 5pt\hbox{$\char'076$}}\mathchar"7218\,$}}}
    \else {${\mathbin{\lower 3pt\hbox
    {$\rlap{\raise 5pt\hbox{$\char'076$}}\mathchar"7218\,$}}}
    $}\fi}
\def\lta{\ifmmode {\,\mathbin{\lower 3pt\hbox   %< or of order
    {$\,\rlap{\raise 5pt\hbox{$\char'074$}}\mathchar"7218\,$}}}
    \else {${\mathbin{\lower 3pt\hbox
    {$\rlap{\raise 5pt\hbox{$\char'074$}}\mathchar"7218\,$}}}
    $}\fi}
\def\lisa{{\it LISA}}

\begin{document}

\title{Binary Encounters With Supermassive Black Holes: Zero-Eccentricity
\lisa\ Events}

\author{M. Coleman Miller\altaffilmark{1}, Marc Freitag\altaffilmark{2},
Douglas P. Hamilton\altaffilmark{1}, and Vanessa M. Lauburg\altaffilmark{1}}

\altaffiltext{1}{Department of Astronomy, University of Maryland at
College Park, College Park, MD 20742-2421; \\
miller@astro.umd.edu, hamilton@astro.umd.edu, vanessa@astro.umd.edu}

\altaffiltext{2}{Department of Physics and Astronomy, Northwestern
University, Evanston, IL 60208; \\
m-freitag@northwestern.edu}

\begin{abstract}

Current simulations of the rate at which stellar-mass compact
objects merge with supermassive black holes (called extreme mass
ratio inspirals, or EMRIs) focus on two-body capture by emission
of gravitational radiation.  The gravitational wave signal of
such events will likely involve a significant eccentricity in the
sensitivity range of the {\it Laser Interferometer Space Antenna}
(\lisa).  We show that tidal separation of stellar-mass compact
object binaries by supermassive black holes will instead produce
events whose eccentricity is nearly zero in the \lisa\ band. 
Compared to two-body capture events, tidal separations have a
high cross section and result in orbits that have a large
pericenter and small apocenter. Therefore, the rate of
interactions per binary is high and the resulting systems are
very unlikely to be perturbed by other stars into nearly radial
plunges.  Depending on the fraction of compact objects that are in
binaries within a few parsecs of the center, the rate of
low-eccentricity \lisa\ events could be comparable to or larger
than the rate of high-eccentricity  events.

\end{abstract}


\vspace{10pt}
\section{Introduction}

Extreme mass ratio inspirals (EMRIs) of stellar-mass compact
objects into supermassive black holes are key targets for the
{\it Laser Interferometer Space Antenna} (\lisa).  From the
fundamental physics standpoint, these events are expected to
provide the best available mapping of the spacetime around a
rotating black hole \citep{Ryan95,Ryan97,Hughes03}. 
Astrophysically, they may well
reveal the numbers of supermassive black holes in a mass range
($\sim 10^5-10^7\,M_\odot$) that is difficult to probe
otherwise \citep[e.g.,][]{GH04}.

Current studies of EMRI rates and properties
\citep{HB95,SR97,MEG00,Freitag01,Freitag03,Ivanov02,HA05}
have focused exclusively on capture of compact objects by
emission of gravitational radiation during a close pass.  That
is, a compact object (for example, a $10\,M_\odot$ black hole)
passes close to the central supermassive black hole (SMBH) and
emits gravitational waves that shrink its orbit
significantly.  The black hole then continues to orbit, and if
its motion is not perturbed significantly by interactions with
other stars then it eventually spirals into the SMBH.  When the
orbit becomes detectable with
\lisa, it has a significant eccentricity of
typically $e\sim 0.5-0.9$ (\citealt{Freitag03,HA05}; but see
\citealt{Ivanov02}).

Here we consider a different process, in which a stellar-mass
binary containing a compact object comes close enough to the SMBH
that the binary is tidally separated, leaving one object bound
to the SMBH and the other almost always ejected to infinity at
high speed.  Tidal separation
was discussed recently by
\citet{Pfahl05} as a way to fuel intermediate-mass black holes,
and
has been considered as a method to produce high-velocity stars
\citep{Hills88,Hills91,YT03,BrownEtAl05} and as a possible
way to deposit high-mass stars close to Sgr A$^*$ \citep{GQ03}.
It was also listed by \citet{HB95} and \citet{FB02b} as a mechanism to be
examined in the EMRI context, but to our knowledge has not yet
been explored quantitatively.

The key point about this process is that, unlike in the
two-body capture scenario, no energy needs to be dissipated in
order to have a capture.  As a result, capture can occur at
much larger radii than is possible in the two-body case: for
example, a binary with a semimajor axis of tenths of an AU can
be captured at pericenter distances of tens of AU relative to
the SMBH,  compared with the $\sim 0.1$~AU that is required
for two-body capture.  In addition, the semimajor axis of the
resulting bound object will be modest, perhaps tens of times
the pericenter distance \citep{Hills91,Pfahl05}.  EMRIs
formed in this way are therefore relatively immune to perturbations
of their orbits that could cause them to plunge directly
into the SMBH (which lowers rates significantly for EMRIs
formed by two-body capture; see \citealt{HB95,HA05}).
Combined with the higher cross section, this suggests that
the overall rate of EMRIs could have an
important contribution from tidal separation of binaries,
even if only a few percent of compact objects are in
binaries.  In addition, the high pericenter after capture
implies that when tidal separation EMRIs are detectable with
\lisa\ they will have eccentricities close to zero \citep[there might
also be independent paths to low eccentricity, such as production
of black holes in an accretion disk and their subsequent advection
to the SMBH; see][]{Levin03}.

In \S~2 we discuss this process in more quantitative detail.
In \S~3 we list some of the questions that will have to be
answered to get more specific predictions of relative rates,
and to interpret \lisa\ observations when they arrive.

\section{Tidal Separation and EMRIs}

\subsection{Capture processes}

To evaluate the tidal separation scenario, let us first recall
the process of two-body capture.  Suppose that a point mass
$m$ orbits a SMBH of mass $M\gg m$ with an orbital speed $v_\infty$ at
apocenter (assumed to be at a large distance).  Its orbit will be
modified significantly if, during its motion, it releases
$\gta {1\over 2}mv_\infty^2$ of energy in gravitational
radiation.  From \citet{QS89}, this condition implies
a pericenter distance
\begin{equation}
r_p<r_{p,{\rm GW}}\approx 0.13~{\rm AU}(m/10\,M_\odot)^{2/7}
(M/10^6\,M_\odot)^{5/7}
(v_\infty/60~{\rm km~s}^{-1})^{-4/7}\; .
\end{equation}
We have scaled by 60~km~s$^{-1}$ because this is roughly the
velocity dispersion inferred for a galaxy with a central black
hole mass of $10^6\,M_\odot$ \citep{MF01,TremaineEtAl02,BGH05}.
The time required to spiral into the SMBH would then be much
less than a Hubble time, except that other stars perturb the
orbit significantly (see \S~2.2).
The gravitational radius is $r_g\equiv GM/c^2\approx 0.01~{\rm
AU}(M/10^6\,M_\odot)$.  Therefore,  
\begin{equation}
r_{p,{\rm GW}}/r_g\approx
13(m/10\,M_\odot)^{2/7}(M/10^6\,M_\odot)^{-2/7}
(v_\infty/60~{\rm km~s}^{-1})^{-4/7}\; .
\end{equation}
For comparison, the radius of the innermost stable circular orbit
around a nonrotating SMBH is $6r_g$.  As another comparison,
detection of an EMRI with \lisa\ will be very difficult if the
gravitational wave frequency is less than $f_{GW}\sim 2-3$~mHz,
because at lower frequencies there is strong unresolvable
foreground noise due to double white dwarf binaries in our Galaxy
\citep{BH97,NYPZ01,FP03}.  For a circular orbit, the gravitational wave
frequency is double the orbital frequency \citep{PM63}.
At 2~mHz, then, the radius of a circular orbit is
$r({\rm 2~mHz})\approx 10\,r_g(M/10^6\,M_\odot)^{-2/3}$.
Therefore, a stellar-mass compact
object needs to go very deep into the potential well of an SMBH
to be captured or to be observed with \lisa.  As a consequence, although
the orbit circularizes due to emission of gravitational radiation
\citep{Peters64}, the eccentricity in the \lisa\ band is still
$e\sim 0.5-0.9$.

Now consider tidal separation.  Suppose that a binary with a
total mass $m$ and semimajor axis $a$ moves towards a
supermassive black hole of mass $M$.  If the orbit has a
pericenter distance less than
\begin{equation}
\begin{array}{rl}
r_{\rm tide}&\approx(3M/m)^{1/3}a\\
&\approx 7~{\rm AU}(M/10^6\,M_\odot)^{1/3}(m/10\,M_\odot)^{-1/3}
(a/0.1~{\rm AU})\; ,
\end{array}
\end{equation}
then the binary will be separated by the tidal field of the SMBH.
Note that the numerical factor in the cube root is correct for a
prograde binary on a circular orbit around the SMBH; it changes
to four for weakly hyperbolic prograde orbits and roughly half
this for retrograde orbits \citep{HB91,HB92}. We scale $a$ by
0.1~AU because such a binary is tight enough to survive
three-body encounters but wide enough to avoid rapid merger by
gravitational radiation (see \S~3 for further discussion).

For an initially hard circular binary with component masses $10\,M_\odot$
and $10\,M_\odot$ in a hyperbolic pass by a $10^6\,M_\odot$ SMBH, our
numerical simulations suggest that the typical eccentricity is $e\sim
0.98$ after capture, consistent with the results of \citet{Hills91} and
\citet{Pfahl05}, who focused on tidal separation of main sequence
binaries.  For an initial binary separation of $a=0.1$~AU, the typical
pericenter distance after capture is a few AU, and the typical apocenter
distance is a few hundred AU; both are proportional to the semimajor axis
of the original binary.  We also simulated tidal separation of initially
hard circular binaries with component masses $10\,M_\odot$ and
$1\,M_\odot$ around a $10^6\,M_\odot$ SMBH, representing for example a
binary with a black hole and a white dwarf or a black hole and a neutron
star.  We find that only a small fraction of encounters lead to ejection
of both objects or survival of the binary, the rest resulting in capture
of one object and ejection of the other.  In $\sim 40\%$
of the captures, the $10\,M_\odot$ object becomes bound to the SMBH, 
with an apocenter distance that is a
factor of a few larger than for the $10\,M_\odot-10\,M_\odot$ simulations
(as is expected given the smaller energy transfer from the $1\,M_\odot$
object; see \citealt{Pfahl05} for an analytic discussion).  In the
remaining $\sim 60\%$ of the captures the $1\,M_\odot$ compact object
is captured, which also leads to an extreme mass ratio inspiral, but a
weaker one than the $10\,M_\odot - 10^6\,M_\odot$ coalescence would
produce.

Because the pericenter distance from binary capture is $\gg r_g$,
the orbit circularizes dramatically by emission of gravitational
radiation and typically has $e<0.01$ in the \lisa\ sensitivity
band, in sharp contrast to EMRIs produced by capture of singles.

\subsection{Effects of nuclear stellar dynamics}

The motion of a binary must
be close to radial to be captured.  For example,
a binary with semimajor axis $a\sim 1$~AU could be captured if
it passed within $\sim 100$~AU of the SMBH, but this is tiny
compared to the distance of a few parsecs from the SMBH where
most binaries presumably lie.  It is therefore important to
map out some of the dynamical processes that will affect the
injection into these orbits.  These are discussed in detail by
many authors \citep[e.g.,][]{FR76,LS77,MT99,SU99}, 
so here we simply quote the results.

A supermassive black hole of mass $M$ will dominate the
dynamics out to the ``radius of influence" 
\begin{equation}
r_{\rm infl}=GM/\sigma_0^2\approx 1~{\rm pc}(M/10^6\,M_\odot)
(60~{\rm km~s}^{-1}/\sigma_0)^2\; ,
\end{equation}
where $\sigma_0$ is the velocity dispersion of stars far outside this radius.
At radii $r>r_{\rm infl}$,
a constant velocity dispersion implies a stellar mass density
$\rho\sim r^{-2}$, whereas at $r<r_{\rm infl}$ the density
can take a different slope $\rho\sim r^{-\gamma}$, for
example $\gamma=3/2$ or $\gamma=7/4$ \citep[e.g.,][]{BW76,Young80}.

For $r<r_{\rm infl}$ the orbital time is
$t_{\rm orb}=2\pi(r^3/GM)^{1/2}$, whereas for $r>r_{\rm infl}$,
$t_{\rm orb}=2\pi (r/r_{\rm infl})(GM/\sigma_0^3)$.  
The relaxation time is the time required for the velocity of a star
to change by of order itself (in magnitude or direction), by
deflections due to two-body encounters.  The local relaxation
time for a compact object of mass $m_{\rm CO}$ interacting with
stars of average mass $\langle m\rangle$ is \citep{Spitzer87}
\begin{equation}
t_{\rm rlx}(r)={0.339\over{\ln\Lambda}}{\sigma^3(r)\over{G^2\langle m\rangle
m_{\rm CO}n(r)}}\; .
\end{equation}
Here $\sigma(r)$ is the local velocity dispersion (equal to the
orbital speed when $r<r_{\rm infl}$), $n(r)$ is the local number
density, and $\ln\Lambda\sim 10$ is the Coulomb logarithm.  
Inside $r_{\rm infl}$ the relaxation time is roughly constant.

For a bound object on a very eccentric orbit, $e\approx 1$, the angular
momentum is much less than the angular momentum of a circular orbit
with the same semimajor axis.  Therefore, the angular momentum only
needs to change slightly to make an order unity difference in the
orbit.  This timescale is 
$t_J(r,e)\approx (1-e)t_{\rm rlx}(r)$ \citep[e.g.,][]{HA05}.

For a given position $R$ and speed $V$, the loss cone is defined as the 
set of directions of the velocity $\vec{V}$
leading to such small pericenter distances that the object of interest is
removed from the system.
In the full loss cone regime, for which $t_J<t_{\rm orb}$ (where $t_J$
is evaluated for the angular momentum corresponding to the loss cone), objects
that enter the loss cone and are removed are immediately replaced,
within an orbital time, by objects that are deflected in from
other orbits.  In this regime, an object that starts down the loss
cone is likely to be deflected out of the cone
during the orbit.  In the empty loss cone regime, for which
$t_J>t_{\rm orb}$, replacement of objects through the loss cone
has to occur over a relaxation time. If the number of objects per 
radius (assuming spherical symmetry)
is $dN/dr$ and the angle subtended by the loss cone at radius $r$
is $\theta_{\rm LC}(r)$, then the approximate capture rates in the
full and empty loss cone regimes are (see \citealt{SU99})
\begin{equation}
\begin{array}{rl}
d{\dot N}_{\rm full}/dr&\sim \theta_{\rm LC}^2(r)(dN/dr)/t_{\rm orb}\\
d{\dot N}_{\rm empty}/dr&\sim (dN/dr)/\left[\ln(1/\theta_{\rm LC}^2(r))
t_{\rm rlx}(r)\right]\; .
\end{array}
\end{equation} 

Far from the SMBH the loss cone is full, whereas close it is empty.
For the binaries, the full/empty transition radius (which defines
the ``critical radius") is comparable to the radius of influence,
whereas for the singles the transition occurs at $\sim 10$\% of the
radius of influence. The merger rate is dominated by the region
near the critical radius (see, e.g., \citealt{FR76}).  The smaller
critical radius for singles partially compensates for the much
larger cross section of the binaries, and for $\gamma=3/2$ the net
merger rate enhancement turns out to be roughly a factor of ten in
favor of the binaries.

As pointed out by \citet{HB95} and analyzed by \citet{HA05},
there is an additional major effect.  A single compact object
captured by gravitational radiation emission typically has a very
large apocenter distance, often on the order of tenths of a
parsec or more.  As a result, even after it has first been
captured, it has a chance to be perturbed in the next orbit.
Sometimes, a perturbation will cause the orbit to be so close to
radial that the object plunges straight into the SMBH.  Although
this does not affect the merger rate, such objects do not
contribute to the \lisa\ event rate, because they plunge before
their orbital period  has become shorter than $\approx
10^{3-4}$~s. \citet{HA05} estimate that $\sim$80-90\% of the
potential EMRI events are lost in this fashion.  This effect
amounts to reducing significantly the volume from which
\lisa-detectable EMRIs can originate, which therefore decreases
the observed rate.  Note, however, that mass segregation of black
holes into a dense subcluster may reduce the impact of this
effect (E.~S. Phinney, personal communication).

In contrast, inspirals produced by separation of binaries are
not susceptible to this effect.  The reason is that, as discussed
in \S~2.1, the apocenter distance is
usually only tens of times the pericenter distance, hence 
$t_J\gg t_{\rm orb}$.  As
a result, we expect that any perturbations will be gradual,
hence a decrease in the pericenter distance will produce
greater gravitational radiation emission and thus circularization
rather than a plunge.

Processes that enhance angular momentum diffusion, such as
resonant relaxation \citep{RT96} and interactions in triaxial potentials
\citep{HolleyBockelManEtAl02,PM02,MP04} will tend to push the
full loss cone regime to smaller radii, which will enhance rates
moderately for both binaries and singles.  However, assessment
of the net effects will require detailed calculations (compare
\citet{RI98}, who show that the total rate of stellar
tidal disruptions is at most doubled by resonant
relaxation, because the rate bottleneck is elsewhere).

\section{Discussion and Conclusions}

In this paper we have focused on interactions that leave a $\sim
10\,M_\odot$ black hole in orbit around a $10^6\,M_\odot$ SMBH.  We
note that the same process will also enhance rates for neutron stars
and white dwarfs around $10^6\,M_\odot$ black holes, likely by an even
larger factor, because we see from Eq. (2) that direct capture of
lower-mass objects at a given speed requires an even closer passage to
the SMBH.   There may also be a moderate effect for intermediate-mass
black holes (IMBHs) of $\sim 10^3\,M_\odot$ in dense stellar clusters
(see \citealt{MC04} for an overview of the evidence for IMBHs and
their association with clusters), as discussed by \citet{Pfahl05}.
This could lead to detection with \lisa\ of BH-IMBH orbits (see
\citealt{BME04} for discussion of gravitational radiation from direct
capture onto IMBHs in clusters) or orbits of disrupted stars around
IMBHs \citep{HPZ05}, although only the nearest sources are likely to
be seen \citep{Will04}. In addition, the loss cone formalism is
unlikely to be directly relevant here because for $M\lta
10^3\,M_\odot$ the wandering radius of the black hole is comparable to
or larger than its radius of influence  \citep[see][]{Merritt01} and
hence the interactions need to be treated as independent binary-single
encounters (see \citealt{GMH04} for a recent application to
intermediate-mass black holes).

It is not trivial to estimate the absolute rate of EMRI captures by
the mechanism we describe, because of the number of processes
involved.  For example, the fraction of black holes in binaries is especially
important (see \citealt{Muno05} for a recent discussion in the
context of the Galactic center).  This fraction depends on (1)~the fraction of
binaries that survive stellar evolution \citep{BSR04}, (2)~the fraction of
those binaries that survive interactions in the dense stellar
environment of the galactic nucleus \citep{IBFR05}, and (3)~the fraction of
initially solitary black holes that acquire companions by three-body
interactions before they are captured by the SMBH.  An important
output from the combination of these processes is (4)~the
distribution of semimajor axes of binaries containing black holes,
because very tight binaries ($a\lta 0.05$~AU for a
$10\,M_\odot-10\,M_\odot$ binary) could merge by gravitational
radiation before being separated by the SMBH, and black holes in
very wide binaries (more than a few AU) could end up after
separation with semimajor axes so large (more than $\sim 0.1$~pc)
that perturbations during a single dynamical time drop them into
plunge orbits, making them undetectable with \lisa\ \citep{HA05}.  Mass
segregation will tend to move black hole binaries to regions of
higher density and higher velocity dispersion, where three-body
interactions are important, hence we must also compute (5)~the
evolution of the semimajor axis and companion mass as a function of
time, versus the probability of capture by the SMBH as a function
of time, to estimate the true distribution of semimajor axes after
capture, and hence the subsequent evolution of the orbit under the
influence of relaxation and gravitational radiation.  All of these
processes will require careful computation in future work.

Without knowing the absolute rate, we can parameterize the
ratio of the rate of captures due to binary separation to the captures
due to singles as
\begin{equation}
{{\dot N}_{\rm binary}\over{{\dot N}_{\rm single}}}=
{f_b {\cal R}_{\rm binary}
\langle f_{\rm binary,LISA}\rangle\over{
f_s{\cal R}_{\rm single}\langle f_{\rm single,LISA}\rangle}}\; ,
\end{equation}
where $f_b$ is the fraction of compact objects that are in
binaries that are neither too tight nor too wide (see above);
$f_s$ is the fraction that are single; ${\cal R}_{\rm binary}$
is the total rate of tidal separations per binary; ${\cal
R}_{\rm single}$ is the total rate of gravitational radiation
captures per single; $\langle f_{\rm binary,LISA}\rangle$ is the
overall fraction of binary sources captured in orbits tight
enough to spiral into the \lisa\ band within a Hubble time; and
$\langle f_{\rm single,LISA}\rangle$ is the overall fraction of
captured singles that end up detectable with \lisa\ (rather than
being perturbed into plunge orbits).  Our current best guesses
are ${\cal R}_{\rm binary}/{\cal R}_{\rm single}\sim 10$ and
$\langle f_{\rm binary,LISA}\rangle / \langle f_{\rm
single,LISA}\rangle\sim 1-10$.   Therefore, if the steady-state
binary fraction (reduced by merging at small semimajor axes, and
by ionizations at large semimajor axes) is $f_b>0.01-0.1$, EMRIs
from binaries could dominate the total rates.

\section*{Acknowledgements}

We thank the Aspen Center for Physics for hospitality and hiking
during the production of this paper, and thank Matt Benacquista, Sam
Finn, Vicky Kalogera, and Alberto Vecchio for organizing the \lisa\  
workshop.  We are grateful to many of our fellow workshop attendees at
Aspen for comments and encouragement.  In particular, Kris Belczynski,
Matt Benacquista, Monica Colpi, Melvyn Davies, Sam Finn, Kelly
Holley-Bockelmann, Vicky Kalogera, Pablo Laguna, Sterl Phinney, Tom
Prince, Fred Rasio, and Dierdre Shoemaker provided many ideas in
discussions, as did Steinn Sigurdsson before the workshop.  We appreciate
the thorough and helpful report of the anonymous referee.  This paper
was supported in part by NASA grant NAG~5-13229. The work of MF at
Northwestern University is funded through NASA ATP grant NAG~5-13236
and his participation in the Aspen center program on \lisa\ data was
supported in part by NASA grant NNG05G106G.

% Here come the reference list (generated with BiBTeX)

%\bibliographystyle{apj}
%\bibliography{biblio} 
% refers to the bibliographic database file ``biblio.bib''

\end{document}